\documentclass[9pt,twocolumn]{article}
\usepackage{lineno,amsmath,graphicx,hyperref}
\usepackage{mathtools}
\usepackage{lmodern}
\usepackage{geometry}
\usepackage{booktabs}
\usepackage{adjustbox}

\makeatletter
\newenvironment{revtexabstract}
{\par\vspace{10pt}\centering%
	\begin{minipage}{0.9\linewidth}%
		\centerline{\large\bfseries Abstract}\vspace{0.5em}\small\noindent\ignorespaces}
	{\end{minipage}\par\vspace{10pt}}
\makeatother

\title{Sub-Poissonian Light in a Waveguide Kerr-medium}

\author{
	R. Singh\textsuperscript{1,}\thanks{ranjit.singh@mail.ru} \\
	\small\textsuperscript{1}Independent Researcher, Domodedovo, 142000, Moscow region, Russia \\
	A.E. Teretenkov\textsuperscript{2} \\
	\small\textsuperscript{2}Department of Mathematical Methods for Quantum Technologies, \\
	\small Steklov Mathematical Institute of Russian Academy of Sciences, \\
	\small 8 Gubkina St., Moscow, 119991, Russia \\
	A.V. Masalov\textsuperscript{3,4} \\
	\small\textsuperscript{3}Russian Quantum Center, Skolkovo Innovation Center, \\
	\small Bolshoi Boulevard 30, 121205, Moscow, Russia \\
	\small\textsuperscript{4}P.N. Lebedev Physical Institute of the Russian Academy of Sciences, \\
	\small Leninsky prosp. 53, 119991 Moscow, Russia
}

\date{\today}

\begin{document}
	
	\maketitle
	
	\begin{revtexabstract}
		Waveguides on a chip represent a new medium for implementing nonlinear optical transformations of light. Modern $\text{Si}_3\text{N}_4$ chip waveguides are attractive due to their high Kerr nonlinearity, suppressed Brillouin scattering on guided acoustic waves (GAWBS), and lengths up to one meter. The capabilities of waveguides for generating sub-Poissonian light in the form of a displaced Kerr state are analyzed. We offer new analytical formulas for estimating the capabilities of suppressing photon fluctuations of a displaced Kerr state for any value of input light amplitude. The results of numerical calculations are presented. It is shown that the degree of photon noise suppression can reach values of 5 - 15 dB with 100 mW light power in waveguides a few meters long.
	\end{revtexabstract}

	\section{Introduction}
	Modern developments in integrated photonics have made possible a new generation of radiation sources in the visible, infrared, and microwave ranges \cite{Gaeta2024,Junqiu2025,Kippenberg2017}. The potential to create new light sources depends on the nonlinear optical properties of the material in on-chip waveguides. Due to the submicron cross-sections of the waveguides and significant lengths, nonlinear properties are developed at continuous radiation powers of 100 mW and less \cite{Tang2019,Brodnik2025,Grange2025}. This enables the implementation of light sources in non-classical quantum states \cite{Vidya2020,Pelucchi2022,Park2024}. On-chip waveguides are competitors with nonlinear optical devices based on optical fibers. Optical fibers have an advantage in nonlinear processes. This is due to their virtually unlimited length. However, fibers demonstrate a serious drawback in the form of Brillouin scattering on guided acoustic waves (GAWBS) for a number of processes based on the Kerr nonlinearity of fiber material (silica) \cite{Shelby1985}. This scattering masks the effects of useful processes. In particular, GAWBS made squeezed light generators on fibers lose the competition with parametric generators on bulk crystals. GAWBS involves oscillations of the fiber cross section that are limited by the side surface, as described in \cite{Poulton2021}. Guided acoustic waves also exist in waveguides as oscillations between two surfaces: the free surface of silica that covers the waveguide and the lower surface of the silicon chip substrate \cite{Thevenaz}. These oscillations pass through the waveguide and interact with light. The intrinsic photoelastic constant of a typical waveguide material, $\text{Si}_3\text{N}_4$, is 6 times smaller than that of the fiber material, $\text{SiO}_2$ \cite{Thevenaz}, which results in a quadratically reduced Brillouin scattering cross section. In addition, there is a decrease in the overlap of the waveguide optical mode and acoustic wave. This decrease is by several times compared to the fiber. This is due to transverse directions. This is demonstrated by computer simulations \cite{Thevenaz}. Consequently, the GAWBS cross section in the waveguide is over two orders of magnitude smaller than that of GAWBS in fibers. This ratio agrees well with the experimental data, where the authors registered Brillouin scattering on a transverse acoustic wave (base of GAWBS) in a pump-probe scheme with counterpropagating beams \cite{Thevenaz}. The reduction of GAWBS cross section in $\text{Si}_3\text{N}_4$ waveguide by almost three orders of magnitude suggests that a number of schemes for nonlinear optical conversion of radiation in waveguides should be reconsidered to build up practical devices. An important limitation of these schemes is the finite length of the waveguides. Currently, realized spiral waveguide lengths are about 1 m on chips measuring a few mm in size \cite{Kippen, Vahala}.
	
	In this paper, we revisit the displaced Kerr state generation scheme to achieve sub-Poissonian photon statistics, adapting it to the capabilities of integrated photonics. We offer new analytic formulas that approximate the values of photon noise suppression at any light power, spectral width, and waveguide length. Our estimations show the ability of modern waveguides on a chip to suppress photon noise. This suppression is comparable to the experimentally demonstrated quadrature noise suppression with squeezed light. That suppression is about 15 dB. (See \cite{Schnabel}.)
	
	Kerr nonlinearity of media is a powerful resource for transforming the properties of light beams and forming new quantum states of radiation. One of the pioneering experiments on generating squeezed light was performed using the Kerr nonlinearity of an optical fiber \cite{Shelby}. Since the first experiment \cite{Shelby}, it has been found that in addition to the Kerr nonlinearity, the generation of squeezed light is accompanied by GAWBS, which adds noise to the propagating radiation and masks the effect of squeezing. To suppress GAWBS, the authors \cite{Shelby} used deep cooling of the fiber. GAWBS was one of the reasons to stop further experiments on the generation of squeezed light with continuous sources in fibers. The development of methods for generating squeezed light due to the Kerr nonlinearity of fibers became possible with pulsed radiation in the soliton regime \cite{Friberg, Spaelter, Schmitt, Krylov, Fiorentino, Tada}.
	
	An alternative to squeezed light is the light with sub-Poissonian photon statistics. The use of such light provides increasing the sensitivity in measuring weak modulation of the radiation intensity and/or weak absorption. In this case, the useful properties of sub-Poissonian light are manifested in schemes with direct photodetection, and there is no need for homodyne detection, which is used for testing and using squeezed light. The latter circumstance is especially convenient in cases where there is no suitable source of a local wave.
	
	The possibility of generating sub-Poissonian light via a displaced Kerr state has been described in theoretical papers \cite{Kitagawa, Knight, Perinova, Ghosh, Sundar, Balybin} and by various mechanisms \cite{Mishra2010a,Mishra2010b,Haderka2017,Mishra2020,Papoff2021}. The main mechanism for suppressing the photon noise of continuous radiation is the Kerr nonlinearity of the medium. Initial coherent light evolves into the "banana" quantum state during interaction with the medium. After appropriate displacement, this state becomes sub-Poissonian light. This state, known as a displaced Kerr state, approaches $N$-photon light. Despite the theoretically confirmed possibility of suppressing photon noise in this scheme, there has been no experimental implementation of sub-Poissonian light generation. At least two reasons can be given for the reduced interest among experimenters in the aforementioned papers \cite{Kitagawa, Knight, Perinova, Ghosh, Sundar, Balybin}. First, the required length of the medium for forming the displaced Kerr state is in the kilometer range. Therefore, experiments can only be planned in fiber-type media. Second, the influence of GAWBS masks the useful effect of the fiber's Kerr nonlinearity.
	
	The formation of quantum properties of light in a Kerr medium has been studied in a number of papers (see \cite{Balybin} and references therein). In theoretical paper \cite{Tanas} it was shown that for input light in coherent state $|\alpha \rangle$ the discrete superpositions of two, three, etc. coherent states arise at substantial interaction lengths. The formation of the superposition of two coherent states, i.e. the Schrödinger cat state, can be achieved when nonlinear phase reaches $\sim |\alpha|^2$. In ordinary solid-state media with  nonlinear refractive index $ n_2 \sim 10^{-19}$ m$^2$/W, kilometer interaction lengths are required. In contrast, the formation of displaced  Kerr state takes place at an early stage of interaction where  the nonlinear phase is about a few radians. At this stage, the quantum state of light takes the form of a "banana" (in the language of Wigner quasiprobability). The center of the "banana" arc is not in the origin of coordinates and the "banana" arc does not cover a full circle. A slight shift of the arc center to the origin makes the "banana" state close to the $N$-photon state. The optimally shifted "banana" state, or displaced Kerr state, becomes sub-poissonian light. Slight shift of quantum state can be achieved in experiments by mixing light with an additional coherent beam on a beam splitter, which is nearly transparent for the beam to be shifted. In this case, the quantum uncertainty body acquires the shift without changing the shape of the quasi-distribution significantly.
	
	In modern literature, the photon noise is usually described by the Mandel parameter $Q = (\langle\Delta \hat{n}^2 \rangle - \langle \hat{n} \rangle ) / \langle \hat{n} \rangle $, which is negative for sub-Poissonian photon statistics and takes a value near $Q=-1$ in the absence of fluctuations as in the $N$-photon state. In the analytical calculation, it is more convenient to describe the photon noise suppression by the value of the Fano factor $F = Q + 1$, which tends to zero as the state approaches to the $N$-photon state.
	
	It should be noted that $N$-photon states are inaccessible in conventional light generation processes due to non-unitary nature of the photon number shift operator. The sub-Poissonian light can serve as a good substitute for $N$-photon radiation and can be useful for generating light in new quantum states.
	
	\section{Theory}
	We study the formation of sub-Poissonian light by considering a single beam in the coherent state at the input to the Kerr-medium
	\begin{eqnarray}
		|\psi (z=0) \rangle =e^{-|\alpha|^2/2} \sum_{n} \frac{\alpha^n}{\sqrt{n!}} |n \rangle. \label{CS}
	\end{eqnarray}
	In the Kerr medium, the quantum state is transformed by the Schrödinger equation with the Hamiltonian \cite{Kitagawa, Tanas}:
	\begin{eqnarray}
		\hat{H} = n_2 \frac{{(\hbar \omega)}^2}{2 n_0 \tau_{\text{coh}} \sigma} \hat{n}^2 = \hbar K v \hat{n}^2, \label{Hint}
	\end{eqnarray}
	and with the evolution operator
	\begin{eqnarray}
		\hat{U} = e^{-it\hat{H}/\hbar} = e^{i K z \hat{n}^2}, \label{Uop}
	\end{eqnarray}
	where $n_0$ is the refractive index of the medium, $v=c/n_0$, $n_2$ is the coefficient of Kerr nonlinearity, $\tau_{\text{coh}}$ is the coherence time of radiation, 
	$\sigma$ is the cross-section of the light beam, and
	\begin{eqnarray}
		K=n_2 \frac{\hbar \omega ^2}{2 c \tau_{\text{coh}} \sigma}.
	\end{eqnarray}
	The correspondence of $t$ and $z$ in (\ref{Uop}) occurs due to the transformation $t\rightarrow t-z/v$.
	
	In papers devoted to the analysis of the quantum Kerr effect, the Hamiltonian may have the form $\hat{H}=\hbar K v \hat{n} (\hat{n}-1)$. The results of quantum calculations with this Hamiltonian do not differ from the case (\ref{Hint}) (see also \cite{Tanas}).
	
	The quantum state of light at the output from the Kerr-medium of length $z$ is given by the expression:
	\begin{eqnarray}
		|\psi (z) \rangle = \hat{U}|\psi(0) \rangle = e^{-|\alpha|^2/2} \sum_{n} e^{i K z n^2}\frac{\alpha^n}{\sqrt{n!}} |n \rangle . \label{CSt}
	\end{eqnarray}
	
\begin{figure}[h]
		\includegraphics[width = \linewidth]{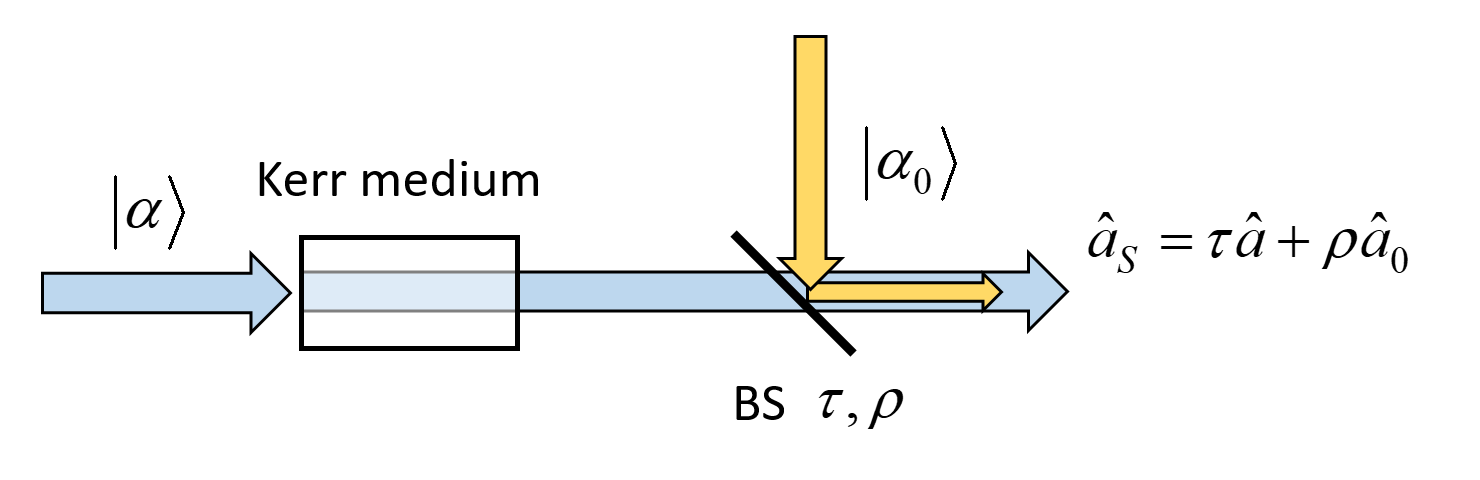}
		\caption{\label{fig:scheme} The scheme of shifting the quantum state of light on a beam splitter BS with additional light $|\alpha_0 \rangle$}
	\end{figure}

	The displacement of the output quantum state can be realized by mixing the light with an additional coherent beam $|\alpha_0 \rangle$ of the same frequency on a beam splitter with almost complete transparency (see Fig.\ref{fig:scheme}). 
	In numerous studies of the shifted Kerr state \cite{Kitagawa, Knight, Perinova, Ghosh, Sundar, Balybin} the mixing was assumed to be realized with the help of the Mach-Zehnder interferometer. In our calculations, we analyze simplified scheme on Fig.\ref{fig:scheme}, which allows selecting the mixing parameters with better flexibility. 
	
	At a small fraction of the added field amplitude $|\rho \alpha_0| \ll 1$, there is a minimal change in the shape of the quasiprobability distribution of the resulting field with small alterations in the mean photon number. The values of the mean photon number and fluctuations of the output state can be calculated by using transformation:
	\begin{eqnarray}
		\hat{a}_S = \tau \hat{a} + \rho \hat{a}_0, \label{modeBS}
	\end{eqnarray}
	where $\hat{a}$ is the operator of the input radiation $|\alpha \rangle$, $\hat{a}_0$ is the operator of the added radiation in the coherent state $|\alpha _0\rangle$, and $\tau$ and $\rho$ are the amplitude coefficients of transmission and reflection of the beam splitter. The Fano factor has the usual form:
	\begin{eqnarray}
		F = \frac{\langle \Delta \hat{n}^2_S \rangle}{\langle \hat{n}_S \rangle} = \frac{\langle \hat{n}^2_S \rangle - \langle \hat{n}_S \rangle^2}{\langle \hat{n}_S \rangle}, \label{FF}
	\end{eqnarray}
	where $\langle \hat{n}_S \rangle = \langle \Psi | \hat{a}^{\dagger}_S \hat{a}_S|\Psi \rangle $ and $\langle \hat{n}_S^2 \rangle = \langle \Psi | \hat{a}^{\dagger}_S \hat{a}_S \hat{a}^{\dagger}_S \hat{a}_S |\Psi \rangle$ with the wave function
	\begin{eqnarray}
		|\Psi \rangle = \sum_{n,m} e^{-|\alpha|^2/2} e^{iKzn^2} \frac{\alpha^n}{\sqrt{n!}} e^{-|\alpha_0|^2/2} \frac{\alpha_0^m}{\sqrt{m!}} |n,m \rangle. \label{psiZ}
	\end{eqnarray}
	The values of $\langle \hat{n}_S \rangle$ and $\langle \Delta \hat{n}^2_S \rangle$ are determined as:
	\begin{align}
		\langle \hat{n}_S \rangle &= \langle \Psi |(\tau ^* \hat{a}^\dagger + \rho ^* \hat{a}^\dagger_0)(\tau \hat{a} + \rho \hat{a}_0)|\Psi \rangle  \nonumber \\
		&= |\tau|^2|\alpha|^2 + \tau^* \rho \alpha_0 \langle \hat{a}^{\dagger} \rangle + \tau \rho^* \alpha_0^* \langle \hat{a} \rangle + |\rho|^2 |\alpha_0|^2 
		\label{Ns}
	\end{align}
	\begin{eqnarray}
		\begin{aligned}
			\langle \Delta \hat{n}^2_S \rangle & = \langle  
			\hat{n}_S^2 \rangle	- \langle \hat{n}_S\rangle^2 \\
			& = 2\tau^{*} |\tau|^2 \rho \alpha_0 (\langle\hat{a}^{\dagger 2} \hat{a}\rangle -|\alpha|^2 \langle \hat{a}^{\dagger} \rangle)\\
			& + 2 \tau |\tau|^2 \rho^* \alpha_0^{*} (\langle\ \hat{a}^{\dagger} \hat{a}^2\rangle -|\alpha|^2 \langle \hat{a} \rangle)\\
			& + \tau^{*2} \rho^2 \alpha_0^2 (\langle \hat{a}^{\dagger 2} \rangle - \langle \hat{a}^{\dagger} \rangle^2) +	\tau^2 \rho^{*2} \alpha_0^{*2} (\langle \hat{a}^2 \rangle - 
			\langle \hat{a} \rangle^2)\\
			& + 2 |\tau|^2 |\rho|^2 |\alpha_0|^2 (|\alpha|^2 - \langle \hat{a}^{\dagger}\rangle \langle \hat{a}\rangle) + \langle \hat{n}_S \rangle,\\
		\end{aligned}   
		\label{varNs}
	\end{eqnarray}
	where it is taken into account that $\langle \hat{a}^{\dagger} \hat{a} \rangle = |\alpha|^2$. In accordance with (\ref{CSt}) for the field averages we have:
	\begin{eqnarray}
		\langle \hat{a} \rangle = \langle \psi| \hat{a} |\psi \rangle = \alpha e^{iKz}e^{2i|\alpha|^2 K z} g_1, 
	\end{eqnarray}
	\begin{eqnarray}
		\langle \hat{a}^2 \rangle = \langle \psi| \hat{a}^2 |\psi \rangle = \alpha^2 e^{4iKz}e^{4i|\alpha|^2 K z} g_2,
	\end{eqnarray}
	\begin{eqnarray}
		\langle \hat{a}^{\dagger}\hat{a}^2 \rangle =  \langle \psi|\hat{a}^{\dagger}\hat{a}^2 |\psi \rangle = \alpha |\alpha|^2 e^{3iKz}e^{2i|\alpha|^2 K z} g_1,
	\end{eqnarray}
	where
	$g_1 = e^{-2i|\alpha|^2 K z} e^{|\alpha|^2(e^{2iK z}-1)}$ and $g_2 = e^{-4i|\alpha|^2 K z} e^{|\alpha|^2(e^{4iK z}-1)}$. The functions $g_1$ and $g_2$ for small $Kz$ are dominated by the real parts. Then, after introducing new notations $\rho \alpha_0 = \alpha_S$ and $\beta = \alpha_S e^{-2i|\alpha|^2Kz}/ (\tau \alpha e^{iK z})$, we have the expression for the Fano factor depending on the normalized variable $\beta$,
	\begin{equation}
		\begin{split}
			F & = \frac{\langle \Delta\hat{n}^2_S \rangle}{\langle \hat{n}_S \rangle} = 1 + \frac{|\tau|^2 |\alpha|^2}{1 + \beta g_1^* + \beta^* g_1 + |\beta|^2}\times \\
			& \times \left[ 2\beta (e^{-2iK z}-1)g_1^* + \text{c.c.} \right.
			+ \beta^2 (e^{-2i K z}g_2^* - g_1^{*2}) + \text{c.c.} \\
			&\quad \left. + 2|\beta|^2 (1 - |g_1|^2) \right] ,
		\end{split}
		\label{Final}
	\end{equation}
	where 
	the value of $\tau$ is close to one.
	Formula (\ref{Final}) is identical to that derived in paper \cite{Sundar} (formulas 40-42), as well as to formulas (4.11, 4.12) of paper \cite{Kitagawa}. We used (\ref{Final}) for numerical search of the minimal Fano factor in complex plane $\beta$ at given values of $\alpha$ and $Kz$.
	
	\section{Numerical analysis}
\begin{figure*}[h]
		\includegraphics[width=\textwidth]{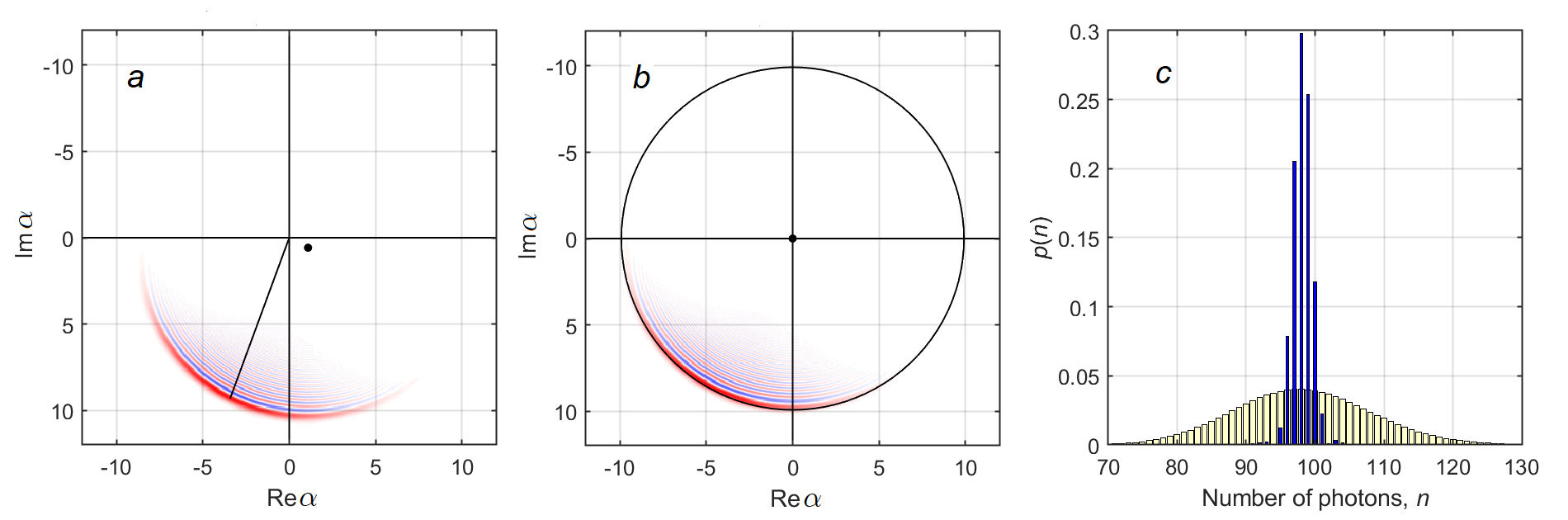}
		\caption{\label{fig:subfig2} Wigner functions: (a) - the state after Kerr medium (\ref{CSt}) at $\alpha$ = 10, black point indicates center of "banana"; (b) - displaced Kerr state. (c) - photon number distribution of the displaced Kerr state (blue) and Poissonian one with the same mean photon number (yellow).}
	\end{figure*}

	The example of numerical calculations is demonstrated in Fig.\ref{fig:subfig2} starting from the Wigner quasiprobability of the light state out of the Kerr medium (Fig.\ref{fig:subfig2}\textit{a}) at the value $\alpha$ = 10 of the input amplitude and optimal length of medium $(Kz)_\text{opt}$ = 0.0218.
	The Wigner quasiprobability was expressed in terms of  coefficients $c_n$ of Fock state decomposition and Laguerre polynomials as follows \cite{Brune}:
	\begin{equation}
		\begin{split}
			&W(\alpha,\alpha^*) = \frac{2}{\pi}\,e^{-2|\alpha|^2}\Big\{ \sum_{n=0}^\infty(-1)^n|c_n|^2L_n(4|\alpha|^2) \\
			&+\sum_{n,m>0}(-1)^n\sqrt{\frac{n!}{(n+m)!}}
			c_{n+m}^*c_n(2\alpha)^m 
			L_n^m(4|\alpha|^2)+\text{c.c.}\Big\}
		\end{split}
		\label{wigfunc}
	\end{equation}
	where $L_n(x)$ and $L^m_n(x)$ are ordinary and associated Laguerre polynomials.
	The black point in the diagram indicates the coordinates of the optimal shift $|\beta |$ = 0.123, which provides the minimum Fano factor. The vector of the optimal shift is approximately perpendicular to the direction of the average phase of the radiation in the Kerr medium $\langle \hat{a} \rangle$ (denoted by the line). The Wigner quasiprobability after the shift (Fig.\ref{fig:subfig2}\textit{b}) demonstrates the transformation of the quantum state into a sub-Poissonian state; the circle in the diagram illustrates the successful choice of shift magnitude. Fig.\ref{fig:subfig2}\textit{c} depicts the resulting photon number distribution (blue amplitudes) with $\langle \Delta \hat{n}^2 \rangle = 1.99$ and $F$ = 0.0203 ($-$16.9 dB); yellow amplitudes illustrate the Poisson distribution with the same mean $\langle \hat{n} \rangle = 98.6$. Significant reduction of photon number fluctuations is evident, accompanied by an insignificant decrease in the mean photon number.
	
	\begin{figure*}[h]
		\includegraphics[width = \linewidth]{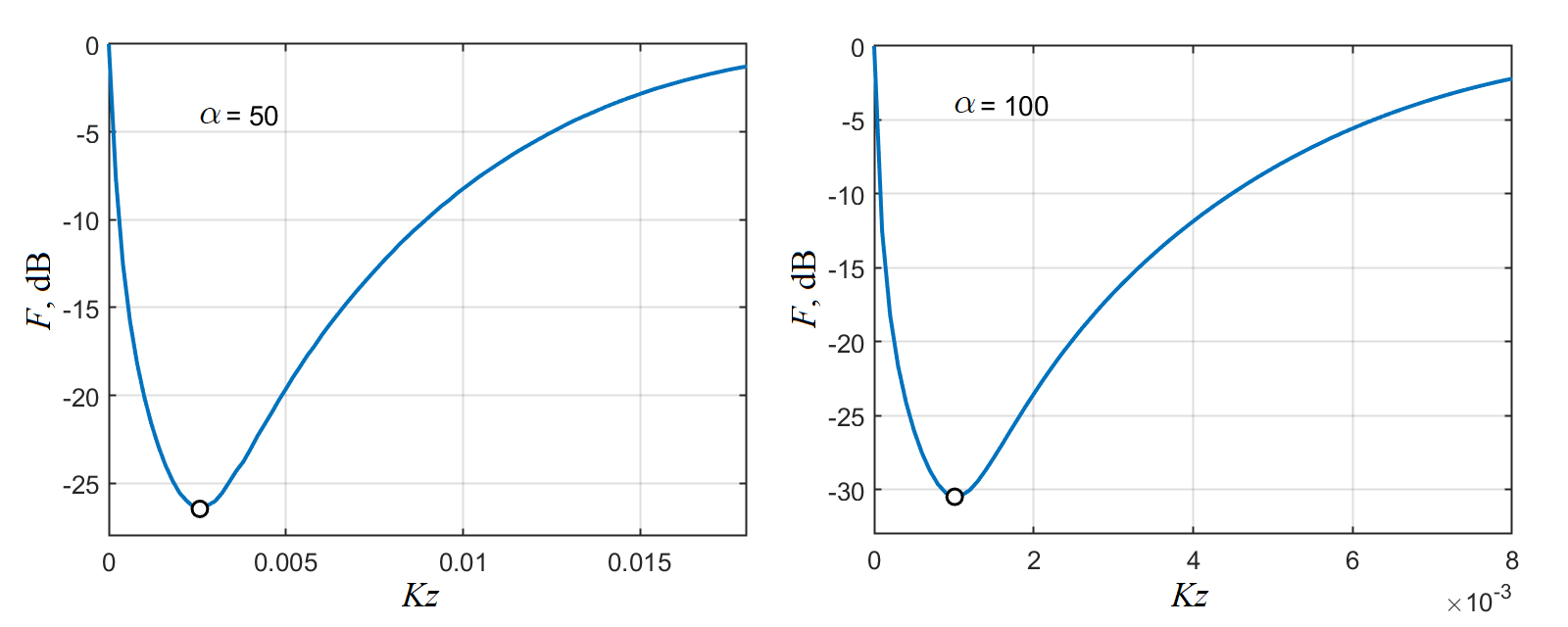}
		\caption{\label{fig:subfig3sh} The dependence of Fano factor of optimally displaced Kerr state on the length of Kerr medium $Kz$ at $\alpha$ = 50 (left) and $\alpha$ = 100 (right); circles denote minimal Fano factors achievable at optimal $Kz_{\text{opt}}$ (see also Table \ref{tab:table1}).}
	\end{figure*}
	Fig.~\ref{fig:subfig3sh} depicts the calculated data on the minimum values of the Fano factor $F$ as a function of the medium length $Kz$ at input state amplitudes $\alpha$ = 50, and 100. All dependencies demonstrate optimal interaction lengths $(Kz)_{\text{opt}}$, which provide the most effective suppression of photon noise. The optimal interaction length was initially identified in paper \cite{Kitagawa}. The calculated data demonstrate that photon noise suppression can reach tens of decibels at relatively low initial light amplitudes. Table \ref{tab:table1} presents the results of the maximum achievable suppression of photon noise $F_{min}$ at optimal medium length $(Kz)_{\text{opt}}$ and added amplitude $|\beta| = |\alpha_s / \alpha|$; the resulting mean photon number is also presented. The added amplitudes, necessary to shift the "banana" state, are relatively small compared to the initial amplitudes. Furthermore, the reduction of mean photon numbers resulting from the shift are negligible.
	
	The results presented in Table \ref{tab:table1} suggest that the high degrees of photon noise suppression are achievable. However, this conclusion is premature. Further estimations will show that the optimal lengths of medium are too long and fall into an unattainable range. 
	To evaluate the realistic potential of photon noise suppression in the regime out of optimal conditions, it is necessary to use analytical formulas for the dependence of the Fano factor on the input amplitudes and lengths of medium, suitable for arbitrary values of variables. Such analytical formulas, which approximate limits of photon noise suppression at $Kz$ from 0 to $\sim (Kz)_{\text{opt}}$ are presented in the next section.
	
	\section{Analytical approximation}
\begin{figure}[h]
		\includegraphics[width = \linewidth]{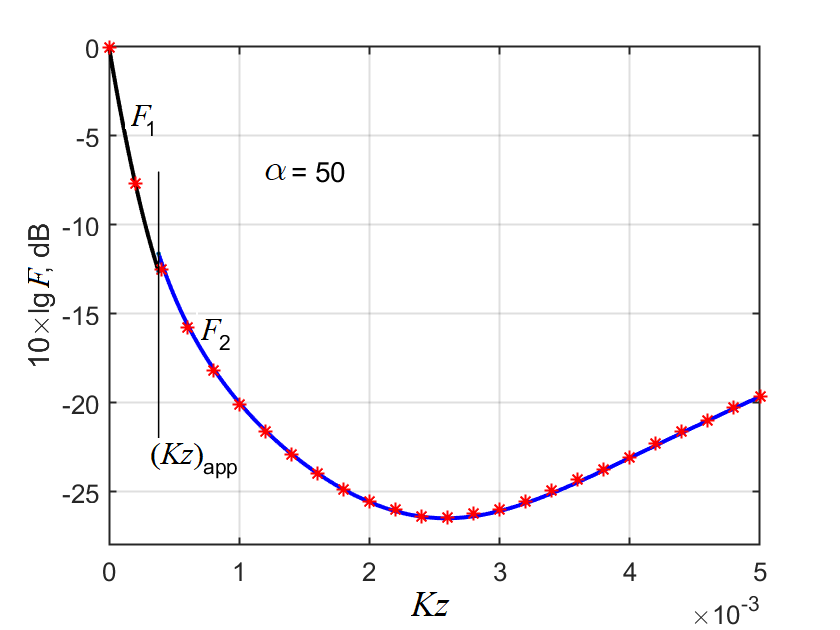}
		\caption{\label{fig:approx} Numerical calculations of the Fano factor of displaced Kerr state versus medium length $Kz$ at $\alpha$ = 50 (red points); black line - approximation (\ref{F1}), blue line - approximation (\ref{F2}).}
	\end{figure}

	The expression (\ref{Final}) for the Fano factor  is the ratio of quadratic polynomials in $\beta$, which satisfies the conditions of the theorem \cite{Tiboulle}. This theorem states that the minimum of such a ratio coincides with the minimum eigenvalue of the matrix forming the polynomials. For the sake of brevity, we omit the rather cumbersome calculations according to this theorem and instead present two resulting approximation formulas for the dependence of the Fano factor on the length of the medium.\\
	\noindent a). Short length approximation:
	\begin{eqnarray}
		F_1 \approx e^{-4|\alpha|^2 K z + |\alpha|^4 (K z)^2} \label{F1}
	\end{eqnarray}
	b). Approximation near optimum lengths:
	\begin{eqnarray}
		F_2 \approx \frac{8}{3}|\alpha|^4 (K z)^4 + \frac{1}{16} \frac{1}{	
			|\alpha|^4 (K z)^2} \label{F2}
	\end{eqnarray}
	These two formulas demonstrate an excellent approximation of numerical results within the range $0\leq Kz \leq 2(Kz)_\text{opt}$ (Fig.\ref{fig:approx}). First formula (\ref{F1}) is applicable at $0\leq Kz \leq (Kz)_\text{app}$, while second one (\ref{F2}) at $(Kz)_\text{app}\leq Kz \leq 2(Kz)_\text{opt}$ where
	\begin{eqnarray}
		(K z)_{\text{app}} = \left(\frac{\sqrt{3}}{2}\right)^{1/3}\frac{1}{|\alpha|^2} \approx \frac{0.953}{	
			|\alpha|^2} .\label{Kzapp}
	\end{eqnarray}
	The photon noise suppression at $(K z)_{\text{app}}$ equals $-$11.6 dB by formula (\ref{Final}) and $-$12.6 dB by formula (\ref{F1}) irrespective of initial amplitude $\alpha$. These two values are close to numerically calculated $-$12.1 dB. Consequently, the suitability of the approximate formulas (\ref{F1}) and (\ref{F2}) can be determined by the noise suppression level. Formula (\ref{F1}) is applicable when $-12.1\leq 10\log F\leq0$ and formula (\ref{F2}) is applicable when $10 \log F \leq -12.1$ dB.
	
	According to (\ref{F2}), the optimal value of $(Kz)_{\text{opt}}$ and the minimum value of the Fano factor $F_{\text{min}}$ can be estimated as:
	\begin{eqnarray}
		(K z)_{\text{opt}} = \frac{1}{2|\alpha|} \left(\frac{\sqrt{3}}{2|\alpha|}\right)^{1/3} \approx \frac{0.477}{|\alpha|^{4/3}}\label{Kzopt}
	\end{eqnarray}
	\begin{eqnarray}
		F_{\text{min}} = \frac{1}{4} \left(\frac{3}{\sqrt{2}|\alpha|^2}\right)^{2/3} \approx \frac{0.413}{|\alpha|^{4/3}}\label{Fmin},
	\end{eqnarray}
	where $(Kz)_{\text{opt}}=(\sqrt{3}/2) F_{\text{min}}$. Approximation $F_{\text{min}}$ by (\ref{Fmin}) coincides with the empirical formula in paper \cite{Perinova}, while authors \cite{Kitagawa} have presented $F_{\text{min}}$ values $\sqrt{3}$ times larger.
	
\begin{figure}[h]
		\includegraphics[width = \linewidth]{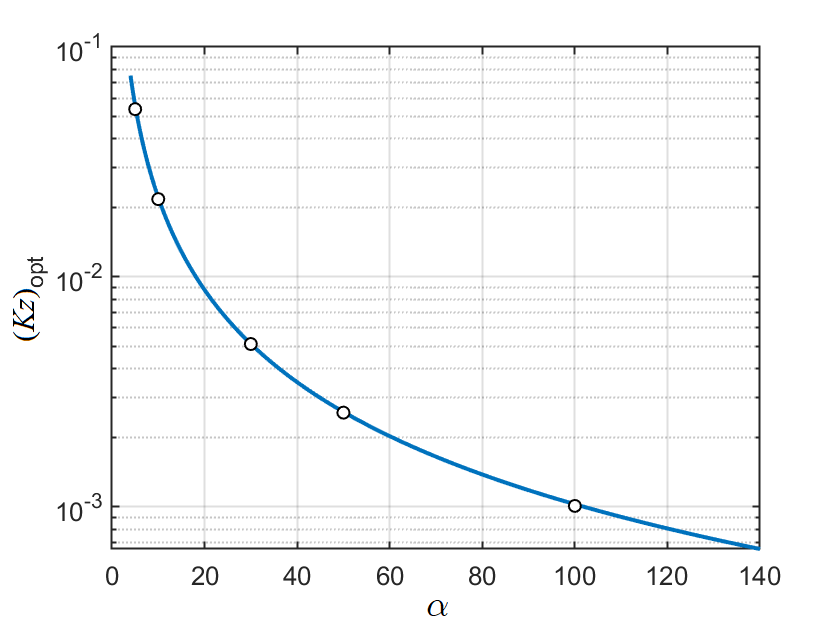}
		\caption{\label{fig:Kzopt} The optimal length of Kerr medium $(Kz)_{\text{min}}$ versus amplitude of initial coherent state $\alpha$: numerical calculations (circles) and approximation (\ref{Kzopt}) (line).}
	\end{figure}

\begin{figure}[h]
		\includegraphics[width = \linewidth]{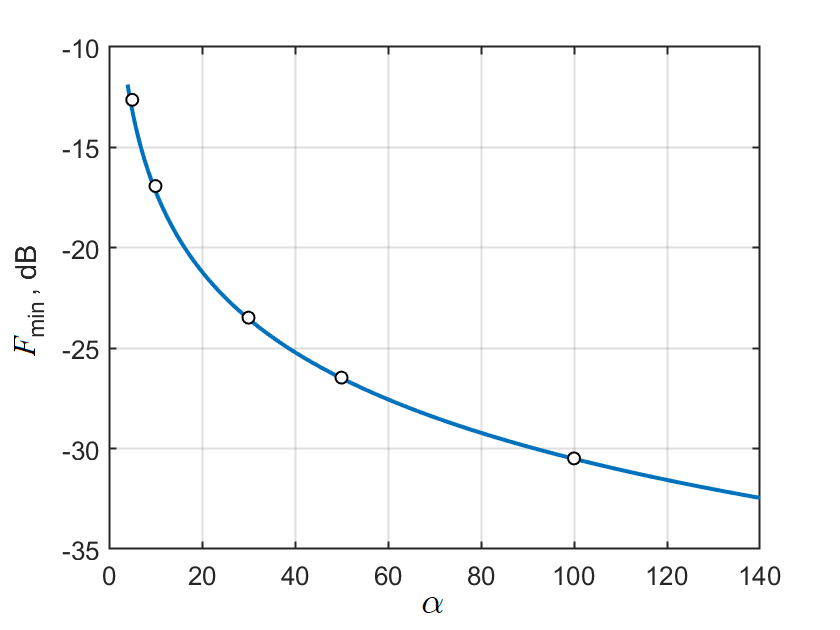}
		\caption{\label{fig:Fmin} The minimal Fano factor $F_{\text{min}}$ achievable with displaced Kerr state versus amplitude of initial coherent state $\alpha$: numerical calculations (circles) and approximation (\ref{Fmin}) (line).}
	\end{figure}

	Approximation formulas (\ref{Kzopt}) and (\ref{Fmin}) demonstrate good agreement with numerical calculations (see Fig.\ref{fig:Kzopt} and Fig.\ref{fig:Fmin})  and data in Table \ref{tab:table1}.
	
	\begin{table}[h]
		\caption{Numerical calculations of optimal noise suppression.}
		\setlength{\tabcolsep}{0.22cm}
		\begin{tabular}{|c|c|c|c|c|}
			\hline
			& & & & \\[-0.3cm]
			$\alpha$ & 10 & 30 & 50 & 100 \\ [0.6ex]
			\hline
			& & & & \\ [-0.3cm]
			$F_{\text{min}}$ & 0.0203 & 0.00449 & 0.00226 & 0.000892 \\
			& -16.9 dB & -23.5 dB & -26.5 dB & -30.5 dB \\ [0.6ex]
			\hline
			& & & & \\ [-0.3cm]
			$(Kz)_{\text{opt}}$ & 0.0218 & 0.00511 & 0.00257 & 0.00102 \\ [0.6ex]
			\hline
			& & & & \\ [-0.3cm]
			$|\beta|=|\alpha_s/\alpha|$ & 0.123 & 0.0569 & 0.0401 & 0.0253 \\ [0.6ex]
			\hline
			& & & & \\ [-0.3cm]
			$\langle n \rangle$ & 98.6 & 894 & 2490 & 9980 \\ [0.6ex]
			\hline
		\end{tabular}
		\label{tab:table1}
	\end{table}
	
	We use (\ref{Kzopt}), (\ref{Fmin}) for estimations of $z_{\text{opt}}$ and $F_{\text{min}}$ in realistic conditions assuming the value of nonlinear phase 
	\begin{eqnarray}
		\varphi _{\text{NL}}= 2 \pi n_2 I z / \lambda = \gamma P z = 2 |\alpha|^2 K z,
	\end{eqnarray}
	where the amplitude $|\alpha|^2$ defines light intensity $I = \hbar \omega |\alpha|^2 / \tau_{\text{coh}} \sigma$ and power $P = \hbar \omega |\alpha|^2 / \tau_{\text{coh}}$ through the width of radiation spectrum $\Delta f = 1/\tau_{\text{coh}}$. The nonlinear parameter $\gamma = 2\pi n_2 / \lambda \sigma_{\text{eff}}$ is accepted in fiber optics. Appearance of the light spectral width in the description of noise suppression is not a surprising fact. Suppression of quantum noise can be attributed only to the number of photons within the quantization volume of the actual mode $c\tau_{\text{coh}}\sigma$. Consequently, the formula for the optimal length of the medium (\ref{Kzopt}) takes the following form:
	\begin{eqnarray}
		\frac{z_{\text{opt}}}{\lambda} = \frac{
			(\sqrt{3}/2)
			^{1/3}}{2 \pi n_2 I} 
		\left(
		\frac{I \tau_{\text{coh}} \sigma}{\hbar \omega}
		\right)^{1/3} 	 = \frac{0.152}{n_2 I}	
		\left(
		\frac{P \tau_{\text{coh}}}{\hbar \omega} \right)^{1/3}.\label{zoptLamb}
	\end{eqnarray}
	
	Table \ref{tab:table2} illustrates the values of optimal medium length $z_{\text{opt}}$ and minimum Fano factor $F_{\text{min}}$ for powers 1 mW, 10 mW and 100 mW and different spectral widths of light with wavelength 1.55 $\mu $m in a $\text{Si}_3\text{N}_4$ waveguide ($n_2 = 2.5 \cdot 10^{-19}$ m$^2$/W) with an effective cross-section $\sigma_{\text{eff}} = 0.3 \cdot 10^{-12}$ m$^2$.
	\begin{table}[h]
		\caption{Values of minimal Fano factor $F_{\text{min}}$ and optimal length of medium $z_{\text{opt}}$ at different values of light power $P$ and spectral width $\Delta f$.}
		\setlength{\tabcolsep}{0.22cm}
		\begin{tabular}{|cc|c|c|c|}
			\hline
			& & & & \\ [-0.3cm]
			$P$ &  & 1 mW & 10 mW & 100 mW \\ [0.6ex]
			\hline
			& & & & \\ [-0.3cm]
			& $\alpha$ & $88\cdot10^3$ & $280\cdot10^3$ & $880\cdot10^3$ \\ [0.6ex]
			$\Delta f=1$ MHz; & $F_{\text{min}}$ & -70 dB & -76 dB & -83 dB \\ [0.6ex]
			& $z_{\text{opt}}$ & 560 km & 120 km & 26 km \\ [0.6ex]
			\hline
			& & & & \\ [-0.3cm]
			& $\alpha$ & $28\cdot10^3$ & $88\cdot10^3$ & $280\cdot10^3$ \\ [0.6ex]
			$\Delta f=10$ MHz; & $F_{\text{min}}$ & -63 dB & -70 dB & -76 dB \\ [0.6ex]
			& $z_{\text{opt}}$ & 260 km & 56 km & 12 km \\ [0.6ex]
			\hline
			& & & & \\ [-0.3cm]
			& $\alpha$ & $8.8\cdot10^3$ & $28\cdot10^3$ & $88\cdot10^3$ \\ [0.6ex]
			$\Delta f=100$ MHz; & $F_{\text{min}}$ & -56 dB & -63 dB & -70 dB \\ [0.6ex]
			& $z_{\text{opt}}$ & 121 km & 26 km & 5.6 km \\ [0.6ex]
			\hline
		\end{tabular}
		\label{tab:table2}
	\end{table}
	
	The data in Table \ref{tab:table2} confirms the previous conclusion that the optimal lengths of medium go beyond the realizable values. Furthermore, the noise suppression degrees of $-$(50-70) dB cannot be measured by modern photocurrent analyzers. Therefore, in our further analysis, it is necessary to refer to estimates of noise suppression at shorter lengths beyond $z_{\text{opt}}$. In such a consideration, the degree of noise suppression will be less and will become amenable to measuring equipment.
	
	Table \ref{tab:table3} presents data on Kerr media lengths that provide the photon noise suppression $-$5, $-$10, and $-$15 dB, estimated by (\ref{F1}) and (\ref{F2}) $F_2 \approx 1/16|\alpha|^4(Kz)^2$ in the same medium and light beam parameters as those in Table \ref{tab:table2}.
	\begin{table}[h]
		\caption{Lengths of Kerr medium required for noise suppression $F$ at light powers $P$ = 10 mW and 100 mW.}
		\setlength{\tabcolsep}{0.32cm}
		\begin{tabular}{|c|c|c|c|}
			\hline
			& & & \\ [-0.3cm]
			$F$ & $|\alpha|^2 Kz$ & $z$ $@P=$10 mW & $z$ $@P=$100 mW \\
			[0.6ex]
			\hline
			& & & \\ [-0.3cm]
			-5 dB & 0.31 & 18 m & 1.8 m \\ [0.6ex]
			\hline
			& & & \\ [-0.3cm]
			-10 dB & 0.70 & 41 m & 4.1 m \\ [0.6ex]
			\hline
			& & & \\ [-0.3cm]
			-15 dB & 1.80 & 82 m & 8.2 m \\ [0.6ex]
			\hline
		\end{tabular}
		\label{tab:table3}
	\end{table}
	
	Data in Table \ref{tab:table3} demonstrate that at least for light power $\geq$ 100 mW the reasonable noise suppression can be experimentally achieved in a waveguide of meter length, available for spiral waveguides on a chip \cite{Kippen, Vahala}.
	
\begin{figure}[h]
		\includegraphics[width = \linewidth]{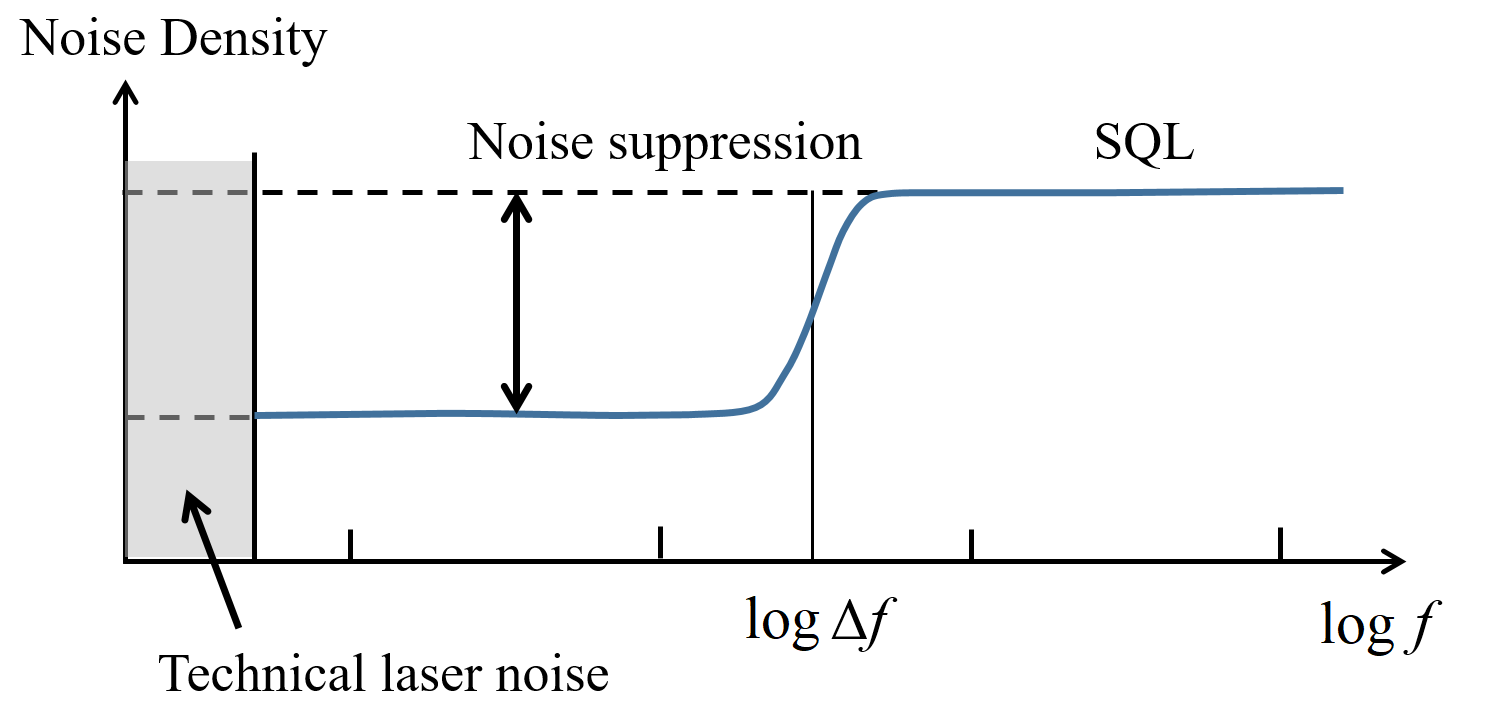}
		\caption{\label{fig:spec} Illustration of photon noise spectral density, SQL - standard quantum level, $\Delta f$ - width of light spectrum.}
	\end{figure}

	Photon noise analysis assumes the usage of direct photodetection and analysis of noise spectral density of photocurrent by an electronic spectrum analyzer. 
	A schematic chart of the noise data is presented in Fig.\ref{fig:spec}. For simplicity it is assumed that the photon noise is detected by a photodiode with unit quantum efficiency and sufficient temporal resolution. The effect of photon noise suppression can be observed only within the frequency range $\Delta f \leq 1/\tau_{\text{coh}}$, while at higher frequencies the spectral density of photocurrent noise power reaches a standard quantum level. At lower frequencies (tens or hundreds of kHz) data on photon noise can be masked by technical laser noise of the initial light beam. By monitoring the spectral density of photocurrent noise power, it is possible to evaluate the width of the spectrum of sub-Poissonian light.
	\begin{figure}[h]
		\includegraphics[width = \linewidth]{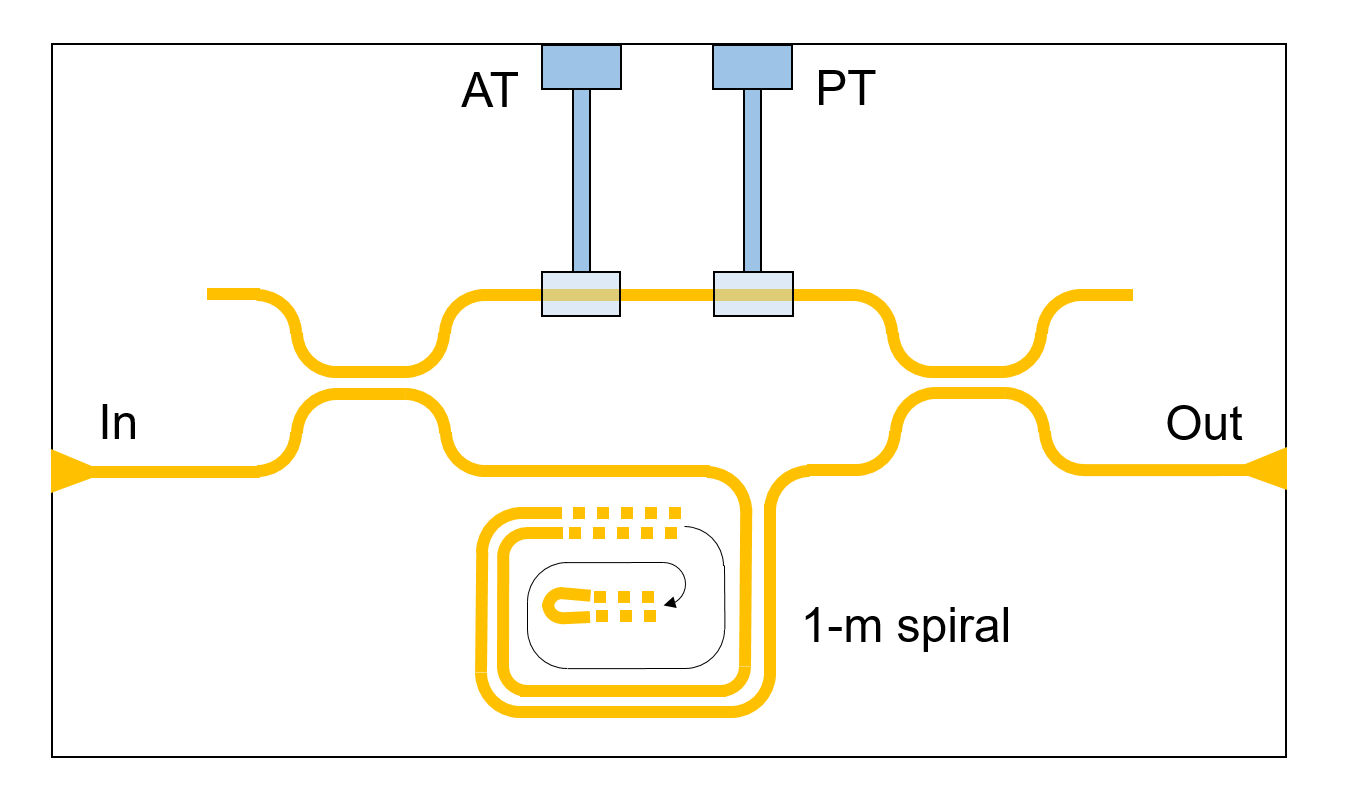}
		\caption{\label{fig:chip} A scheme of a waveguide Mach-Zehnder interferometer on a chip for generating a displaced Kerr state is presented. The AT electrode controls the amplitude of the transmitted radiation (amplitude tuning). PT controls the phase (phase tuning).}
	\end{figure}
	
	\section{Conclusion}
	
	The data on Brillouin scattering on a guided acoustic wave \cite{Thevenaz} allow for the expectation of an insignificant influence of GAWBS when forming a displaced Kerr state in on-chip waveguides. Our theoretical analysis of the photon noise of the displaced Kerr state provides analytical formulas for the calculation of the degree of noise suppression for arbitrary values of radiation power, spectrum width, and waveguide length. Although there is an optimum in photon noise suppression for the length of the medium, this optimum is unattainable in practice due to the requirement for an unrealistically long waveguide length. Estimates show that in a realistic situation, when the waveguide length is less than the optimum, significant photon noise suppression can nevertheless be achieved. It is shown that the degree of photon noise suppression can reach values of 5 - 15 dB with 100 mW light power in waveguides a few meters long.
	The scheme for displaced Kerr state generation is usually based on a Mach-Zehnder interferometer (see, for example, \cite{Knight}). This scheme in the form of waveguides on a chip could have the form shown in Fig. \ref{fig:chip}.
	
	\section*{Funding}
	A.V.M. acknowledges financial support from the Russian Science Foundation (Project No. 23-42-00111).
	
	\section*{Acknowledgment}
	A.V.M. is also grateful to M. Vasilyev for insightful comments. 
	
	\section*{Disclosures}
	The authors declare no conflicts of interest.
	
	\section*{Data availability} Data underlying the results presented in this paper are not publicly available at this time but may be obtained from the authors upon reasonable request.

\end{document}